\documentclass[12pt]{article}
\usepackage{amssymb,amsmath,epsfig}
\allowdisplaybreaks

\begin{document}

\title{\bf Stellar Structures admitting Noether Symmetries in $f(\mathcal{R},\mathcal{T})$
Gravity}
\author{M. Sharif \thanks {msharif.math@pu.edu.pk} and
M. Zeeshan Gul \thanks{mzeeshangul.math@gmail.com}\\
Department of Mathematics, University of the Punjab,\\
Quaid-e-Azam Campus, Lahore-54590, Pakistan.}

\date{}
\maketitle

\begin{abstract}
This paper investigates the geometry of compact stellar objects via
Noether symmetry strategy in the framework of curvature-matter
coupled gravity. For this purpose, we assume the specific model of
this theory to evaluate Noether equations, symmetry generators and
corresponding conserved parameters. We use conserved parameters to
examine some fascinating attributes of the compact objects for
suitable values of the model parameters. It is analyzed that compact
objects in this theory depend on the conserved quantities and model
parameters. We find that the obtained solutions provide the
viability of this process as they are compatible with the
astrophysical data.
\end{abstract}
\textbf{Keywords:} $f(\mathcal{R},\mathcal{T})$ theory; Compact stellar
structures; Noether symmetries; Conserved quantities.\\
\textbf{PACS:} 04.20.Jb; 04.50.Kd; 98.35.Ac; 98.80.Jk

\section{Introduction}

The current cosmic acceleration has been the most stunning and
dazzling consequence for the scientific community over the last two
decades. Scientists claim that this expansion is the outcome of some
cryptic force named dark energy which has antigravitational effects.
This ambiguous force has inspired many scientists to uncover its
hidden aspects. In this regard, alternative gravitational theories
are assumed as the most elegant and significant approaches to reveal
the dark universe. These proposals can be established by modifying
the geometric and matter part of the Einstein-Hilbert action. The
simplest modification of general relativity (GR) is $f(\mathcal{R})$
theory which is established by inserting the function of Ricci
scalar $(\mathcal{R})$ in the geometric part of the generic action.
To comprehend the viability of this theory, a significant literature
has been made available in \cite{6}.

The $f(\mathcal{R})$ gravity has further been generalized by
establishing some coupling between curvature and matter parts. These
couplings explain the rotation curves of galaxies and different
cosmic stages. These proposals are non-conserved that yield the
existence of an extra force. These interaction proposals are
extremely helpful to comprehend mysterious aspects of the
cosmos.\cite{7}. Harko et al. \cite{9} constructed these couplings
in $f(\mathcal{R})$ theory named $f(\mathcal{R},\mathcal{T})$
gravity. The non-minimal coupling of geometry with matter was
formulated in \cite{10}, dubbed as
$f(\mathcal{R},\mathcal{T},\mathcal{R}_{\alpha\beta}
T^{\alpha\beta})$ theory. Another interaction provides
$f(\mathcal{R}, \mathcal{T}^{\phi})$ theory \cite{11}. The
metric-affine $f(\mathcal{R},\mathcal{T})$ theories of gravity and
their applications has been discussed in \cite{11a}.

Symmetry plays a crucial role in the study of cosmology and
gravitational physics. Accordingly, the Noether symmetry (NS)
technique is considered the most efficient approach that describes a
correlation between symmetry generators and conserved parameters of
a physical system. This method minimizes the complexity of the
system and generates new solutions which are then discussed in terms
of cosmic features. These symmetries are not just a method to deal
with the dynamical solutions but their presence can also provide
some viable conditions so that cosmological models can be selected
according to current observations \cite{10a}. The NS approach is
also used to investigate the nature of dark components \cite{10b}.
Moreover, this technique is an important and systematic way to
evaluate the conserved parameters. Conservation laws play an
important role in studying different physical phenomena. These laws
are the specific cases of the Noether theorem which states that any
differentiable symmetry of the action corresponds to some
conservation law. This theorem is most significant because it gives
information about the conservation laws in physical theories. The
conservation laws of linear and angular momentum determine the
translational and rotational symmetry of any object. The Noether
charges are important in the literature as they are used to examine
some important cosmological problems in different contexts.
\cite{10c}.

In modified gravitational theories, the NS approach has many
significant applications. Capozziello et al. \cite{12} established
the analytic solutions of spherical metric in $f(\mathcal{R})$
theory admitting NS methodology. The exact cosmological solutions by
NS technique in $f(\mathcal{R},\mathcal{G})$ theory has been found
in \cite{13}. The NS of FRW universe for different matter
configuration in $f(\mathcal{G})$ theory has been studied in
\cite{14}. Shamir and Ahmad \cite{15} investigate various
cosmological models through NS approach in
$f(\mathcal{G},\mathcal{T})$ theory. Sharif et al \cite{16} used
this strategy in a different context to analyze the dark universe.
Recently, we have obtained exact cosmological solutions in
energy-momentum squared gravity and analyzed their behavior through
various physical parameters \cite{16a}. We have also investigated
wormhole solutions and geometry of compact stellar objects in this
background \cite{16b}.

Researchers are fascinated by the properties and consequences of
celestial objects due to their interesting characteristics and
relativistic structures in cosmology and astrophysics. Gravitational
collapse is the result of this event and it is responsible for the
production of new dense stars known as compact objects. Because of
their large masses and small radii, these objects are thought to be
extremely dense. Alternative gravitational theories and GR
adequately characterize these compact objects \cite{17}. Abbas et
al. \cite{18} examined the physical aspects and stable state of
compact objects in $f(\mathcal{G})$ theory. The structure of
anisotropic compact objects in $f(\mathcal{R})$ theory has been
examined in \cite{19}. The geometry of compact objects through NS
strategy in the modified theory has been investigated in \cite{20}.
The effect of alternative gravitational theories is familiar to
examine the structure of compact objects and fluid configuration
\cite{21}-\cite{24}.

This article examines symmetry generators and associated conserved
parameters for a specific $f(\mathcal{R},\mathcal{T})$ model. We
then analyze the most important characteristics of compact objects
for various values of model variables. The manuscript is planned as
follows. Section \textbf{2} establishes the equations of motion of
spherical spacetime in $f(\mathcal{R},\mathcal{T})$ theory. Section
\textbf{3} examines the NS technique. In section \textbf{4}, we use
conserved parameters and appropriate conditions to identify the
expression of metric elements. Section \textbf{5} analyzes physical
attributes of the compact stars to investigate the viability of the
model through graphical interpretation. A brief summary and
discussion of the results are given in the last section.

\section{Basic Formalism of $f(\mathcal{R},\mathcal{T})$ Gravity}

The action of this theory is determined as \cite{9}
\begin{equation}\label{1}
S= \int{\left(\frac{f(\mathcal{R},\mathcal{T})}{2\kappa^2}+
L_{m}\right)\sqrt{-g}d^4x},
\end{equation}
where $g$ and $\kappa^2$ determine the determinant of the metric
tensor and coupling constant, respectively. We assume $\kappa^2$ as
a unity for our convenience. The following equations of motion are
formulated by varying the action corresponding to the metric tensor
\begin{equation}\label{2}
\mathcal{R}_{\alpha\beta}f_{\mathcal{R}}+g_{\alpha\beta}\Box
f_{\mathcal{R}}-\nabla_{\alpha}\nabla_{\beta}f_{\mathcal{R}}
-\frac{1}{2}g_{\alpha\beta}f =T_{\alpha\beta}-(T_{\alpha\beta}
+\Theta_{\alpha\beta})f_{\mathcal{T}},
\end{equation}
where $\Box= \nabla_{\alpha}\nabla^{\alpha}$, $f\equiv
f(\mathcal{R}, \mathcal{T})$, $f_{\mathcal{T}}=
\frac{\partial f}{\partial \mathcal{T}}$, $f_{\mathcal{R}}
= \frac{\partial f}{\partial \mathcal{R}}$ and
\begin{eqnarray}\label{3}
\Theta_{\alpha\beta}=
g_{\alpha\beta}L_{m}-2T_{\alpha\beta}-2g^{\vartheta\eta}\frac
{\partial^{2}L_{m}}{\partial g^{\alpha\beta}\partial g^{\vartheta\eta}}.
\end{eqnarray}
It is noted that, this theory takes the form of $f(\mathcal{R})$
theory for $f(\mathcal{R},\mathcal{T}) = f(\mathcal{R})$ and reduces
to GR when $f(\mathcal{R},\mathcal{T}) =\mathcal{R}$. The
stress-energy tensor demonstrates the matter configuration in
gravitational physics and provides dynamical parameters with some
physical characteristics.

We consider matter configuration as an isotropic fluid
\begin{equation}\label{4}
T^{m}_{\alpha\beta}=
\emph{U}_{\alpha}\emph{U}_{\beta}\left(\mathrm{p}_{m}
+\mathrm{\rho}_{m}\right)+g_{\alpha\beta}\mathrm{p}_{m},
\end{equation}
where energy density, pressure and  four velocity of the matter is
represented by , and $\mathrm{\rho}_{m}$, $\mathrm{p}_{m}$ and
$\emph{U}_{\alpha}$, respectively. Manipulating Eq.(\ref{3}), we get
\begin{eqnarray}\nonumber
\Theta_{\alpha\beta}=
-2T^{m}_{\alpha\beta}+\mathrm{p}_{m}g_{\alpha\beta}.
\end{eqnarray}
By using Eq.(\ref{2}), we have
\begin{equation}\label{5}
G_{\alpha\beta}=
\frac{1}{f_{\mathcal{R}}}\left(T_{\alpha\beta}^{c}+T_{\alpha\beta}^{m}
\right)=T_{\alpha\beta}^{eff},
\end{equation}
where $T_{\alpha\beta}^{c}$ are the additional effects of modified
theory and $T_{\alpha\beta} ^{eff}$ determines the effective
energy-momentum tensor defined as
\begin{eqnarray}\nonumber
T_{\alpha\beta}^{eff}&=&
\frac{1}{f_{\mathcal{R}}}\bigg\{\left(1+f_{\mathcal{T}}\right)
T_{\alpha\beta}^{m}+\left(\nabla_{\alpha}\nabla_{\beta}-g_{\alpha
\beta}\Box\right)f_{\mathcal{R}}-\mathrm{p}_{m}g_{\alpha\beta}f_{\mathcal{T}}
\\\label{6}&+&
\frac{1}{2}g_{\alpha\beta}\left(f-\mathcal{R}f_{\mathcal{R}}\right)\bigg\}.
\end{eqnarray}
We take into account static spherical spacetime to examine the
characteristics of compact objects \cite{25}.
\begin{equation}\label{7}
ds^{2}=
-e^{a\left(\emph{r}\right)}dt^{2}+e^{b\left(\emph{r}\right)}dr^{2}
+\emph{r}^{2}\left(d\theta^{2}+\sin^{2}\theta d\phi^{2}\right).
\end{equation}
The respective field equations are
\begin{eqnarray}\nonumber
\mathrm{\rho}^{eff}&=&
\frac{1}{f_{\mathcal{R}}}\left\{\left(1+f_{\mathcal{T}}\right)
\mathrm{\rho}_{m}-\frac{1}{2}\left(f-\mathcal{R}f_{\mathcal{R}}
\right)+\mathrm{p}_{m}f_{\mathcal{T}}+e^{-b}\left\{f_{\mathcal{R}}''
\right.\right.\\\label{8}&-&\left.\left.
\left(\frac{a'}{2}-\frac{2}{r}\right)f_{\mathcal{R}}'\right\}\right\},
\\\nonumber
\mathrm{p}^{eff}&=&
\frac{1}{f_{\mathcal{R}}}\left\{\left(1+f_{\mathcal{T}}\right)
\mathrm{p}_{m}+\frac{1}{2}\left(f-\mathcal{R}f_{\mathcal{R}}\right)
-e^{-b}\left(\frac{a'}{2}+\frac{2}{r}\right)f_{\mathcal{R}}'
\right.\\\label{9}&-&\left.
\mathrm{p}_{m}f_{\mathcal{T}}\right\}.
\end{eqnarray}
The field equations (\ref{8}) and (\ref{9}) are very complicated due
to the inclusion of multivariate function and their derivatives. The
direct solution of these equations is very difficult. There are two
possible ways to solve these equations. One is to solve them by
applying a suitable exact or numeric method whereas 2nd is to obtain
exact solutions by NS technique. As this theory is non-conserved but
we achieved conserved parameters through the NS approach which are
then used to analyze the geometry of compact objects. Thus the
latter approach appears to be more interesting and we adopt it in
this article.

\section{Point-like Lagrangian and Noether Symmetry}

Here, we develop the point-like Lagrangian for static spherical
spacetime in the context of $f(\mathcal{R},\mathcal{T})$ theory. The
canonical form of the action (\ref{1}) is determined as
\begin{equation}\label{10}
S=
\int \mathcal{L}\left(a,b,\mathcal{R},\mathcal{T},a',b',\mathcal{R}'
,\mathcal{T}'\right)dr.
\end{equation}
To obtain point-like Lagrangian, we use the Lagrange multiplier
technique as
\begin{equation}\label{11}
S=
\int \sqrt{-g}\Big\{f-(\mathcal{R}-\bar{\mathcal{R}})\mu_{1}-
\left(\mathcal{T}-\bar{\mathcal{T}}\right)\mu_{2}+\mathrm{p}_{m}
\left(a,b\right)\Big\}dr,
\end{equation}
where
\begin{eqnarray}\nonumber
\sqrt{-g}=e^{\frac{a+b}{2}}\emph{r}^{2}, \quad \mathcal{\bar{T}}
=3\mathrm{p}_{m}-\mathcal{\rho}_{m}, \quad \mu_{1}=f_{\mathcal{R}},
\quad \mu_{2}=f_{\mathcal{T}},
\\\label{12}
\bar{\mathcal{R}}=-\frac{1}{e^{b}}\Big(a''+\frac{a'^{2}}{2}+\frac{2a'}
{\emph{r}}-\frac{2b'}{\emph{r}}-\frac{a'b'}{2}-\frac{2e^{b}}{\emph{r}
^{2}}+\frac{2}{\emph{r}^{2}}\Big).
\end{eqnarray}
We observe that the action (\ref{11}) reduces to action (\ref{1})
when $\mathcal{R}-\bar{\mathcal{R}}=0$ and $\mathcal{T}-
\bar{\mathcal{T}}=0$. The corresponding Lagrangian turns out to be
\begin{eqnarray}\nonumber
&&\mathcal{L}\left(a,b,\mathcal{R},\mathcal{T},a',b',\mathcal{R}',\mathcal{T}'\right)
=\emph{r}^{2}e^{\frac{a+b}{2}}\Big(f+\mathrm{p}_{m}-\mathcal{R}f_{\mathcal{R}}+\frac
{2f_{\mathcal{R}}}{\emph{r}^2}
\\\nonumber
&&+f_{\mathcal{T}}\left(3\mathrm{p}_{m}-\mathrm{\rho}_{m}-\mathcal{T}\right)\Big)
+\emph{r}^{2}e^{\frac{a-b}{2}}\bigg\{\left(\frac{2b'}{\emph{r}}-\frac{2}{\emph{r}^{2}}
\right) f_{\mathcal{R}}+a'\mathcal{R}'f_{\mathcal{R}\mathcal{R}}
\\\label{13}
&&+a'\mathcal{T}'f_{\mathcal{R}\mathcal{T}}\bigg\}.
\end{eqnarray}
The corresponding Euler-Lagrange equations ($\frac{\partial
\mathcal{L}}{\partial q^{i}}-\frac{d}{dr}\left(\frac{\partial
\mathcal{L}} {\partial q^{i'}}\right)=0$) give
\begin{eqnarray}\nonumber
&&f-\mathcal{R}f_{\mathcal{R}}+\mathrm{p}_{m}+f_{\mathcal{T}}\left(3\mathrm{p}_{m}
-\mathrm{\rho}_{m}+6\mathrm{p}_{m_{,a}}-2\mathrm{\rho}_{m_{,a}} -\mathcal{T}\right)
+2\mathrm{p}_{m_{,a}}
\\\nonumber
&& +\frac{1}{e^{b}}\Bigg\{-2\mathcal{R}'^{2}f_{\mathcal{R}\mathcal{R}\mathcal{R}}
-4\mathcal{R}'\mathcal{T}'f_{\mathcal{R}\mathcal{R}\mathcal{T}}+\left(\frac{2b'}
{\emph{r}}+\frac{2e^{b}}{\emph{r}^{2}}-\frac{2}{\emph{r}^{2}}\right)f_{\mathcal{R}}
\\\nonumber&&
+\left(b'\mathcal{R}'-2\mathcal{R}''-\frac{4\mathcal{R}'}{\emph{r}}\right)f_{\mathcal{R}
\mathcal{R}}+\left(b'\mathcal{T}'-2\mathcal{T}''-\frac{4\mathcal{T}'}{r}\right)
f_{\mathcal{R}\mathcal{T}}
\\\label{15}&&
-2\mathcal{T}'f_{\mathcal{R}\mathcal{T}\mathcal{T}}\Bigg\}=0,
\\\nonumber&&
f-\mathcal{R}f_{\mathcal{R}}+\mathrm{p}_{m}+f_{\mathcal{T}}\left(3\mathrm{p}_{m}
-\mathrm{\rho}_{m}+6\mathrm{p}_{m_{,b}}-2\mathrm{\rho}_{m_{,b}}-\mathcal{T}\right)
+2\mathrm{p}_{m_{,b}}
\\\nonumber&&
+\frac{1}{e^{b}}\Bigg\{\left(\frac{2e^{b}}{\emph{r}^{2}}-\frac{2a'}{\emph{r}}-\frac{2}
{\emph{r}^{2}}\right)f_{\mathcal{R}}-\left(a'\mathcal{R}'+\frac{4\mathcal{R}'}{\emph{r}}
\right)f_{\mathcal{R}\mathcal{R}}-a'\mathcal{T}'f_{\mathcal{R}\mathcal{T}}
\\\label{16}&&
-\frac{4\mathcal{T}'}{r}f_{\mathcal{R}\mathcal{T}}\Bigg\}=0,
\\\nonumber&&
e^{b}\bigg\{\left(\mathcal{R}-\frac{2}{\emph{r}^{2}}\right)f_{\mathcal{R}\mathcal{R}}
-\left(3\mathrm{p}_{m}-\mathrm{\rho}_{m}-\mathcal{T}\right)f_{\mathcal{R}\mathcal{T}}
\bigg\}+\left(a''+\frac{a'^{2}}{2}
\right.\\\label{17}&&\left.
+\frac{2a'}{\emph{r}}-\frac{2b'}{\emph{r}}-\frac{a'b'}{2}+\frac{2}{\emph{r}^{2}}\right)
f_{\mathcal{R}\mathcal{R}}=0,
\\\nonumber&&
e^{b}\bigg\{\left(\mathcal{R}-\frac{2}{\emph{r}^{2}}\right)f_{\mathcal{R}\mathcal{T}}
-\left(3\mathrm{p}_{m}-\mathrm{\rho}_{m}-\mathcal{T}\right)f_{\mathcal{T}\mathcal{T}}
\bigg\}+\left(a''+\frac{a'^{2}}{2}
\right.\\\label{18}&&\left.
+\frac{2a'}{\emph{r}}-\frac{2b'}{\emph{r}}-\frac{a'b'}{2}+\frac{2}{\emph{r}^{2}}\right)
f_{\mathcal{R}\mathcal{T}}=0.
\end{eqnarray}

The Hamiltonian is determined as
\begin{eqnarray}\nonumber
H&=&
-e^{\frac{a-b}{2}}\emph{r}^{2}\Big\{e^{b}\left(f+\mathrm{p}_{m}+\left(3\mathrm
{p}_{m}-\mathrm{\rho}_{m}-\mathcal{T}\right)f_{\mathcal{T}}-\mathcal{R}f_{\mathcal{R}}
\right.\\\label{20}&+&\left.
2\emph{r}^{-2}f_{\mathcal{R}}\right)-2\emph{r}^{-2}f_{\mathcal{R}}-a'\mathcal{R}'
f_{\mathcal{R}\mathcal{R}}-a'\mathcal{T}'f_{\mathcal{R}\mathcal{T}}\Big\}.
\end{eqnarray}
The symmetry generators are given by
\begin{equation}\label{21}
Y= \varrho(a, b,\mathcal{R}, \mathcal{T})\frac{\partial}{\partial
\emph{r}}+\xi(a, b,\mathcal{R},
\mathcal{T})^{i}\frac{\partial}{\partial q^{i}},
\end{equation}
where undetermined coefficients of $Y$ are represented by $\varrho$
and $\xi$, respectively. For the presence of Noether symmetries, the
Lagrangian must fulfill the invariance condition defined as
\begin{equation}\label{22}
Y^{[1]}\mathcal{L}+(D\varrho)\mathcal{L}= D\psi,
\end{equation}
where $D$ demonstrates the total rate of change, $Y^{[1]}$ defines
the first order prolongation and $\psi$ is the boundary term. This
can also be defined as
\begin{equation}\label{23}
Y^{[1]}= Y+{\xi^{i}}'\frac{\partial}{\partial{q^{i}}'},
~~~D=\frac{\partial} {\partial
\emph{r}}+{q^{i}}'\frac{\partial}{\partial {q^{i}}},
\end{equation}
here ${\xi^{i}}'= D{\xi^{i}}'-{q^{i}}'D\varrho$. The corresponding
conserved quantities are defined as
\begin{equation}\label{24}
I=
-\varrho H+ \xi^{i}\frac{\partial \mathcal{L}}{\partial q^{i}}-\psi.
\end{equation}
This is also dubbed as the first integral of motion and considered
the most crucial part of NS which play a significant role to
determine viable cosmological solutions and  determine the
attributes of compact objects in modified theories. We have the
following system of equations by comparing the coefficients of Eq.
(\ref{22}).
\begin{eqnarray}\label{25}
&&\varrho_{,a}=0, \quad \varrho_{,b}=0, \quad \varrho_{,\mathcal{R}}=0,
\quad \varrho_{,\mathcal{T}}=0,
\\\label{26}
&&a_{,b}=0, \quad a_{,\mathcal{R}}=0, \quad a_{,\mathcal{T}}=0,
\\\label{27}
&&\xi^{3}_{,b}f_{\mathcal{R}\mathcal{R}}+\xi^{4}_{,b}f_{\mathcal{R}
\mathcal{T}}=0, \quad \xi^{3}_{,a}f_{\mathcal{R}\mathcal{R}}+\xi^{4}
_{,a}f_{\mathcal{R}\mathcal{T}}=0,
\\\label{28}
&&r^{2}\xi^{1}_{,\emph{r}}f_{\mathcal{R}\mathcal{R}}+2\emph{r}\xi^{2}
_{,\mathcal{R}}f_{\mathcal{R}}-e^{\frac{b-a}{2}}\psi_{,\mathcal{R}}=0,
\\\label{29}
&&r^{2}\xi^{1}_{,\emph{r}}f_{\mathcal{R}\mathcal{T}}+2\emph{r}\xi^{2}
_{,\mathcal{T}}f_{\mathcal{R}}-e^{\frac{b-a}{2}}\psi_{,\mathcal{T}}=0,
\\\label{30}
&&2\emph{r}\xi^{2}_{,a}f_{\mathcal{R}}+\emph{r}^{2}\xi^{3}_{,\emph{r}}
f_{\mathcal{R}\mathcal{R}}+r^{2}\xi^{4}_{,\emph{r}}f_{\mathcal{R}
\mathcal{T}}-e^{\frac{b-a}{2}}\psi_{,a}=0,
\\\label{31}
&&\left(a-b+2\xi^{2}_{,b}\right)\emph{r}f_{\mathcal{R}}+2\emph{r}\xi^{3}
f_{\mathcal{R}\mathcal{R}}+2\emph{r}\xi^{4}f_{\mathcal{R}\mathcal{T}}-e^
{\frac{b-a}{2}}\psi_{,b}=0,
\\\nonumber
&&2\emph{r}^{2}\xi^{3}f_{\mathcal{R}\mathcal{R}\mathcal{R}}+2\emph{r}^{2}
\xi^{4}f_{\mathcal{R}\mathcal{R}\mathcal{T}}+2\emph{r}^{2}\xi^{4}
_{,\mathcal{R}}f_{\mathcal{R}\mathcal{T}}
\\\label{32}
&&+\left(a-b+2\xi^{1}_{,a}+2\xi^{3}_{,\mathcal{R}}-2\varrho_{,\emph{r}}
\right)r^{2}f_{\mathcal{R}\mathcal{R}}=0,
\\\nonumber
&&2\emph{r}^{2}\xi^{3}f_{\mathcal{R}\mathcal{R}\mathcal{T}}+2\emph{r}^{2}
\xi^{4}f_{\mathcal{R}\mathcal{T}\mathcal{T}}+2\emph{r}^{2}\xi^{3}
_{,\mathcal{T}}f_{\mathcal{R}\mathcal{R}}
\\\label{33}
&&+\left(a-b+2\xi^{1}_{,a}+2\xi^{4}_{,\mathcal{T}}-2\varrho_{,\emph{r}}
\right)\emph{r}^{2}f_{\mathcal{R}\mathcal{T}}=0,
\\\nonumber
&&e^{\frac{a+b}{2}}r^{2}\bigg\{\left(f+\mathrm{p}_{m}-\left(\mathcal{R}
-2\emph{r}^{-2}\right)f_{\mathcal{R}}+\left(3\mathrm{p}_{m}-\mathrm{\rho}
_{m}-\mathcal{T}\right)f_{\mathcal{T}}\right)
\\\nonumber
&&\times\frac{1}{2}\left(\xi^{1}+\xi^{2}+2\varrho_{,\emph{r}}\right)+\xi^{1}
\left(\mathrm{p}_{m_{,a}}+\left(3\mathrm{p}_{m_{,a}}-\mathrm{\rho}_{m_{,a}}
\right)f_{\mathcal{T}}\right)
\\\nonumber
&&+\xi^{2}\left(\mathrm{p}_{m_{,b}}+\left(3\mathrm{p}_{m_{,b}}-\mathrm{\rho}
_{m_{,b}}\right)f_{\mathcal{T}}\right)-\xi^{3}\left(\left(\mathcal{R}-2\emph{r}
^{-2}\right)f_{\mathcal{R}\mathcal{R}}
\right.\\\nonumber&&\left.
+\left(3\mathrm{p}_{m}-\mathrm{\rho}_{m}-\mathcal{T}\right)f_{\mathcal{R}
\mathcal{T}}\right)-\xi^{4}\left(\left(3\mathrm{p}_{m}-\mathrm{\rho}_{m}
-\mathcal{T}\right)f_{\mathcal{T}\mathcal{T}}
\right.\\\label{34}&&\left.
+\left(\mathcal{R}-2\emph{r}^{-2}\right)f_{\mathcal{R}\mathcal{T}}\right)
+\frac{2f_{\mathcal{R}}}{\emph{r}e^{b}}\psi_{,\emph{r}}\bigg\}-\psi_{,\emph{r}}=0.
\end{eqnarray}
The NS technique minimizes the system's complexity and helps to
derive the analytic solutions. Nonetheless, its difficult to
formulate viable solutions without considering any particular
$f(\mathcal{R},\mathcal{T})$ model. In the curvature-matter
interaction, the study of compact objects using the NS strategy
would yield fascinating results. To examine the geometry of compact
objects, we examine the presence of symmetry generators and
conserved parameters.

In the present cosmic era, it is analyzed that several celestial
objects lie in the nonlinear phase. To obtain a veritable picture of
their structural formation and evolution, we have to look at their
linear behavior. Therefore, we take into account a minimal model
defined as \cite{9}
\begin{eqnarray}\label{34a}
f(\mathcal{R},\mathcal{T})=
f_{1}\left(\mathcal{R}\right)+f_{2}\left(\mathcal{T}\right).
\end{eqnarray}
This cosmological model can include an appropriate extension of
$f(\mathcal{R})$ theory. The feasible $f(\mathcal{R},\mathcal{T})$
gravity models can be established from Eq.(\ref{34a}) by considering
distinct forms of $f_{1}\left(\mathcal{R}\right)$ and $f_{2}\left(
\mathcal{T}\right)$. In the present work, we take into account
$f_{1} \left(\mathcal{R}\right)= \mathcal{R}+\mu \mathcal{R}^{2}$
known as Starobinsky model \cite{26} and
$f_{2}\left(\mathcal{T}\right)= \nu \mathcal{T}$, where $\mu$ and
$\nu$ are constants \cite{27}. In this regard, the
$f(\mathcal{R},\mathcal{T})$ gravity model (\ref{34a}) turns out to
be
\begin{eqnarray}\label{34b}
f(\mathcal{R},\mathcal{T})=
\mathcal{R}+\mu \mathcal{R}^{2}+\nu \mathcal{T}.
\end{eqnarray}
This model efficiently explains the radiation-dominated era and
considered as an alternative candidate for dark energy \cite{28}.
For $\nu=0$, this model reduces to the Starobinsky model and GR is
recovered when $\mu=0=\nu$. In this paper, we take into account the
above model (\ref{34b}) to examine the geometry of compacts objects.
This model has been deeply discussed in the literature to study the
geometry of compact objects and collapsing phenomenon \cite{28a}.

A dust fluid can also explain the exact matter configuration of
different celestial objects. Here, we analyze the attributes of
compact objects and formulate analytic solutions for dust matter
distribution, i.e.,
$T^{m}_{\alpha\beta}=\mathcal{\rho}_{m}\emph{U}_{\alpha}\emph{U}_{\beta}$.
The exact solutions of Eqs.(\ref{25})-(\ref{34}) yield
\begin{eqnarray}\nonumber
&&\varrho= c_{3}\emph{r}, \quad \xi^{1}=c_{1}, \quad \xi^{2} =
c_{1}+2c_{2}, \quad \xi^{3}= \frac{2c_{2}(1+\mathcal{R} \mu)}{\mu},
\\\nonumber
&&\mathrm{\rho}_{m}= \frac{1}{\left(c_{1}+c_{2}+c_{3}\right)
\nu\emph{r}^2}\{(2c_{1}+4c_{2}+2c_{3})(1+2\mu\mathcal{R})
\\\nonumber
&&-(c_{1}+3c_{2}+c_{3})\mu\emph{r}^2\mathcal{R}^{2}-c_{2}
\emph{r}^2\mathcal{R}\}, \quad \xi^{4}= 0= \psi,
\end{eqnarray}
where $c_{i}$ represent the arbitrary constants. The symmetry
generators and associated conserved quantities turn out to be
\begin{eqnarray}\nonumber
&&Y_{1}
=\frac{\partial}{\partial a}+\frac{\partial}{\partial b}, \quad
Y_{2}
= \frac{\partial}{\partial b}+\mathcal{R}\frac{\partial}
{\partial \mathcal{R}}, \quad Y_{3}= r\frac{\partial}{\partial r},
\\\label{35}
&&I_{1}= 2\emph{r}e^{\frac{a-b}{2}}\left(1+2\mu\mathcal{R}+\mu
\emph{r}\mathcal{R}'\right),~ I_{2}= 2\emph{r}e^{\frac{a-b}{2}}
(1+2\mu\mathcal{R})(4+a'\emph{r}),
\\\label{36}
&&I_{3}= -e^{2}\emph{r}\{2+4\mu\mathcal{R}+2\mu\emph{r}^{2}
a'\mathcal{R}'-\left(2+4\mu\mathcal{R}-\mu\emph{r}^{2}\mathcal
{R}^{2}\right)e^{b}\}.
\end{eqnarray}

\section{Metric Elements and Boundary Conditions}

The conserved parameters derived using the NS strategy are useful
for studying a variety of physical properties of compact objects.
This method has been studied in both axial/spherical spacetimes
\cite{29}. The smooth matching of inner and outer geometries is used
to investigate the solutions at the surface boundary. The metric
elements are linked as
\begin{equation}\label{37}
e^{a(\emph{r})}= e^{-b(\emph{r})}.
\end{equation}
We take into account the above conserved quantities (\ref{36}) and
(\ref{37}) to formulate the metric elements that provide useful
consequences to examine the viable characteristics of compact
objects. Using Eq.(\ref{37}) in (\ref{35}) and (\ref{36}), we have
\begin{eqnarray}\label{39}
&&I_{1}=
2\emph{r}e^{a}\left(1+2\mu\mathcal{R}+\mu\emph{r}\mathcal{R}'\right), \quad
I_{2}=
2\emph{r}e^{a}(1+2\mu\mathcal{R})(4+a'\emph{r}),
\\\label{40}
&&I_{3}=
\emph{r}\{2+4\mu\mathcal{R}-\mu\emph{r}^{2}\mathcal{R}^{2}
-(2+4\mu\mathcal{R}+2\mu\emph{r}^{2}a'\mathcal{R}')e^{a}\}.
\end{eqnarray}
We can not solve the above equations analytically as these are in
complex forms. Therefore, we use a numeric approach to analyze the
viable behavior of metric potentials.

The metric elements inside the compact objects must be non-singular,
monotonically increasing and regular to analyze the physically
realistic cosmological model. The graphical representation of metric
potentials obtained by $I_{1}$, $I_{2}$ and $I_{3}$ are shown in
Figures (\textbf{1}, \textbf{2}, \textbf{3}), respectively, which
show that metric potentials satisfy all the required conditions. The
following conditions must be fulfilled for physically realistic
compact stars.
\begin{itemize}
\item  The effective fluid parameters must be positive inside the
stellar geometry as well as on the surface boundary whereas pressure
should be zero at the surface boundary.
\item  The gradient of effective matter variables should
be negative for $0\leq r\leq \mathrm{R}$, i.e.,
$\bigg\{\left(\frac{d\mathrm{\rho}^{eff}}{d\emph{r}}\right)
_{\emph{r}=0}=0$ , $\left(\frac{d^{2}\mathrm{\rho}^{eff}}
{d\emph{r}^{2}}\right)_{\emph{r}=0}<0$ ,
$\left(\frac{d\mathrm{p}^{eff}}{d\emph{r}}\right) _{\emph{r}=0}=0$
and $\left(\frac{d^{2}\mathrm{p}
^{eff}}{d\emph{r}^{2}}\right)_{\emph{r}=0}<0\bigg\}$. This condition
determines that the effective fluid variables must be decreasing at
the boundary of surface.
\item  The speed of light must be greater than sound speed.
\end{itemize}
The geometry of compact objects is determined by these physical
features. Here, we study the physical characteristics of celestial
objects for metric potentials attained by $I_{1}$ as the fluid
parameters become undefined for metric elements obtained by the
$I_{2}$ while $I_{3}$ is in the complex form and we can not obtain
the suitable value of metric potentials.
\begin{figure}\center
\epsfig{file=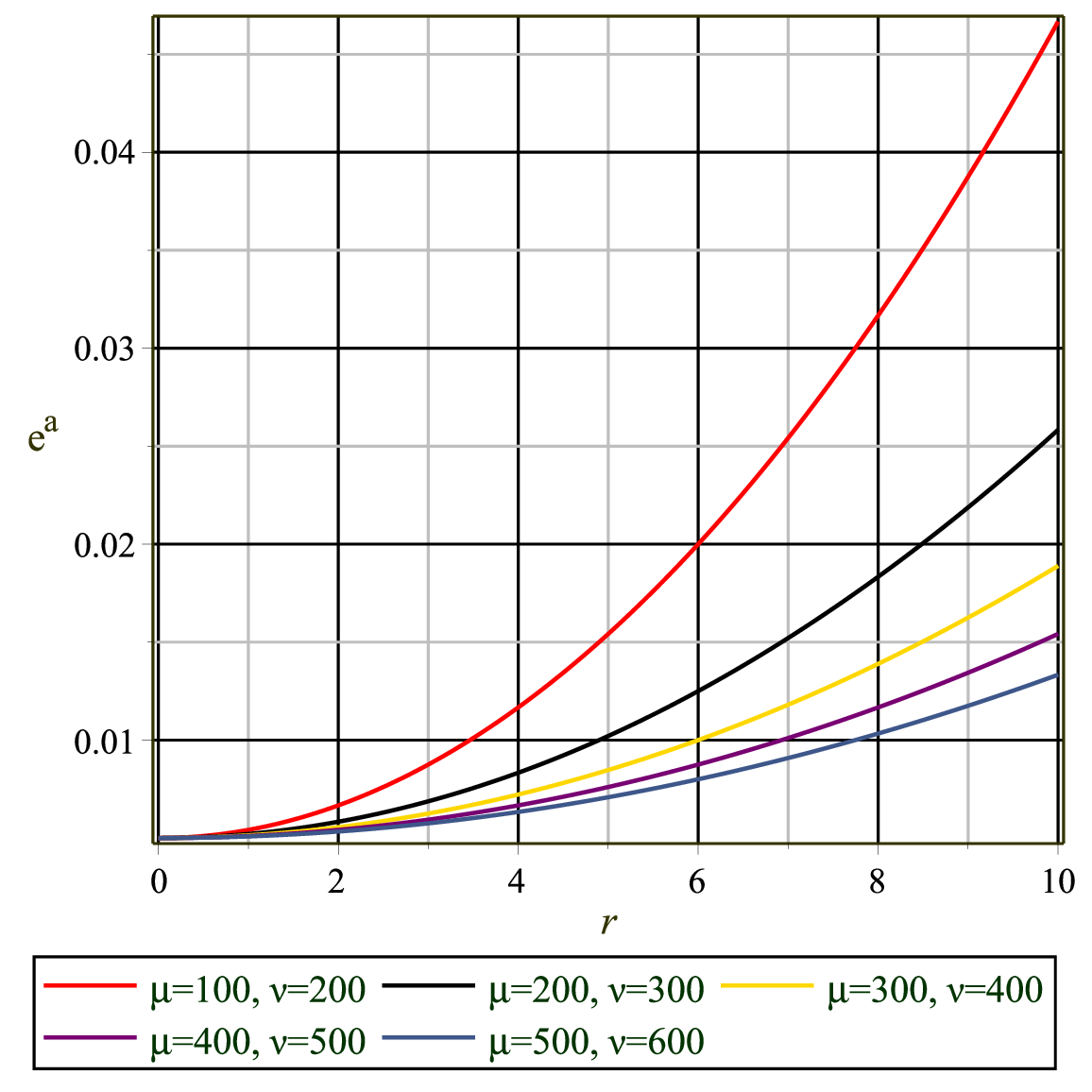,width=.5\linewidth} \caption{Graph of metric
potential versus $\emph{r}$.}
\end{figure}
\begin{figure}\center
\epsfig{file=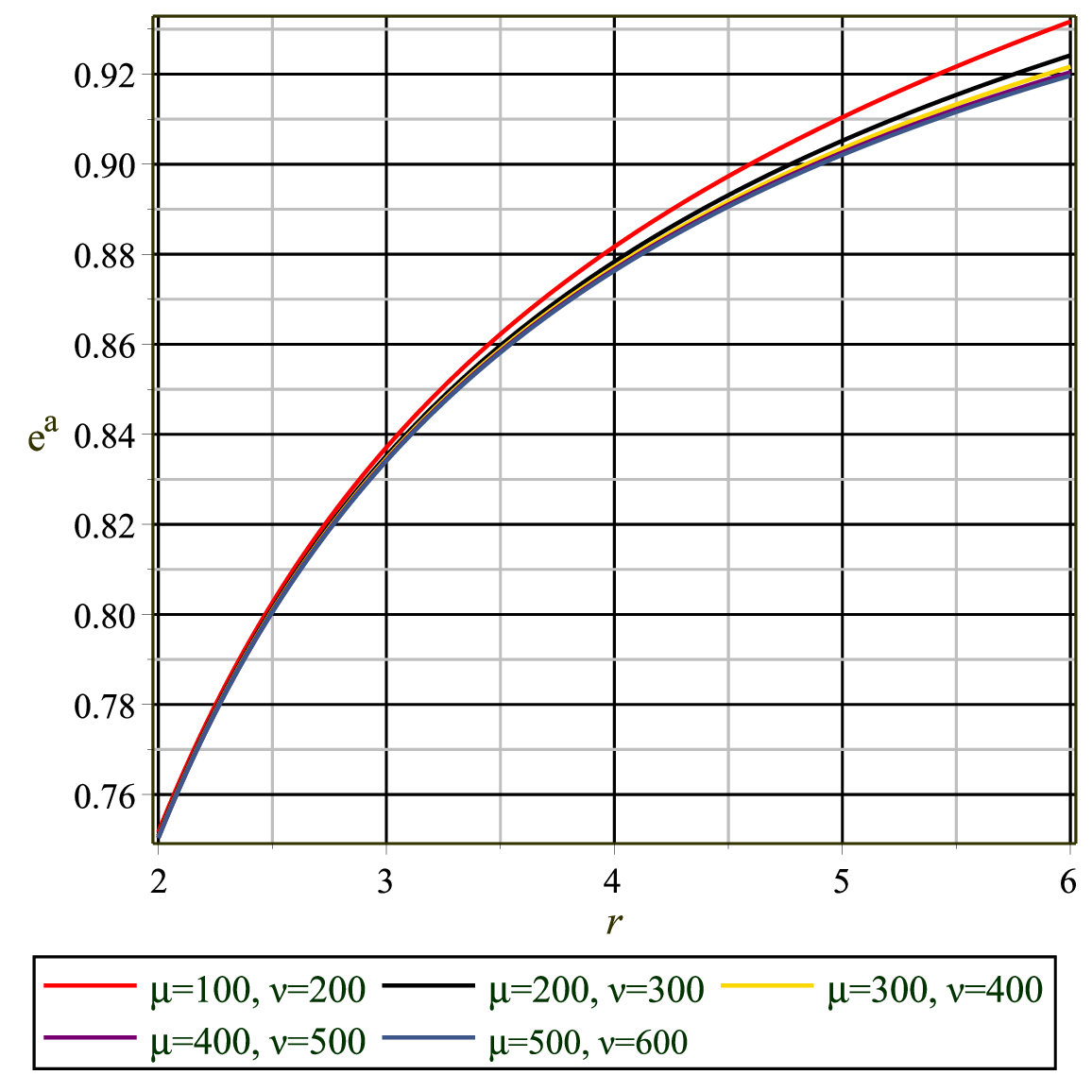,width=.5\linewidth} \caption{Graph of metric
potential versus $\emph{r}$.}
\end{figure}
\begin{figure}\center
\epsfig{file=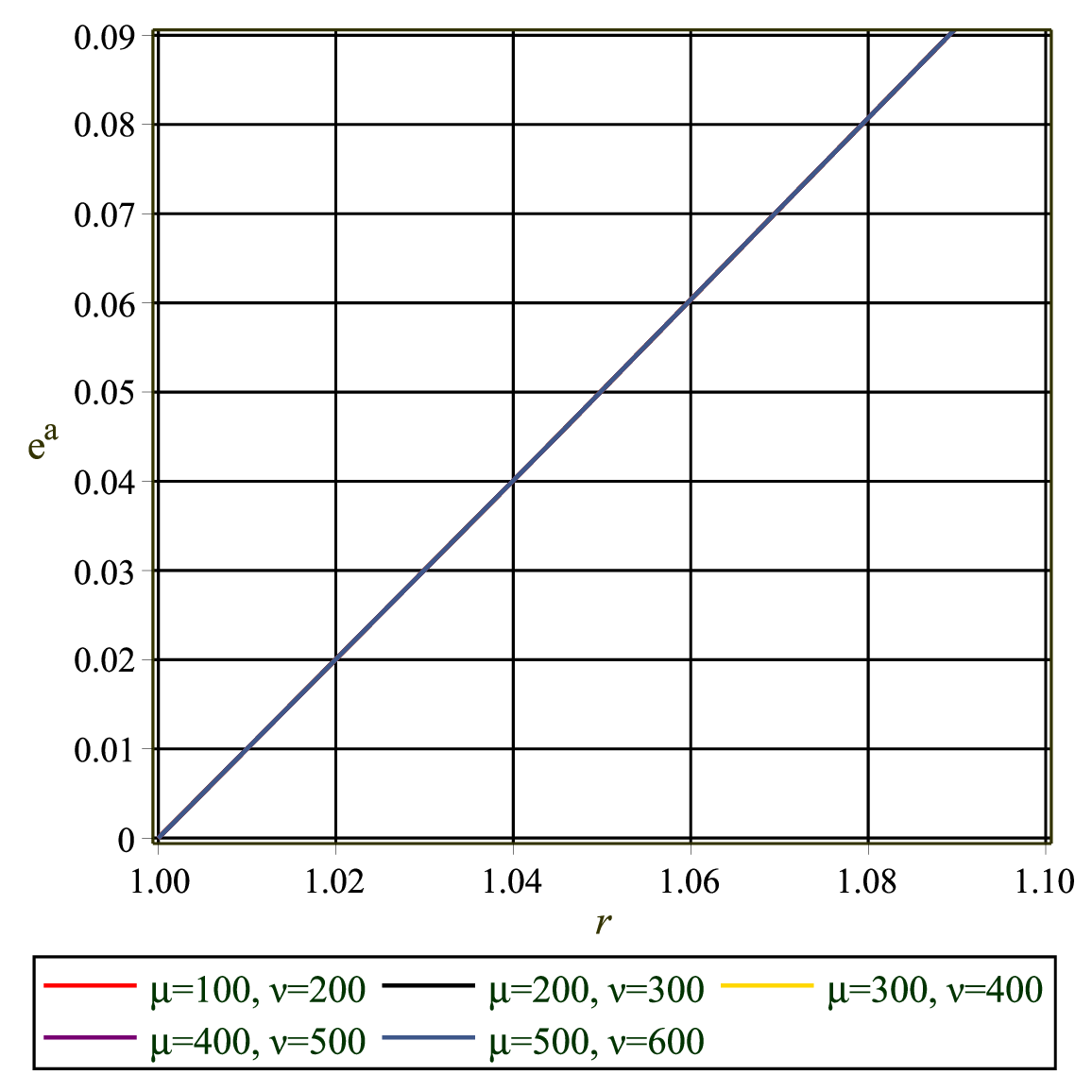,width=.5\linewidth} \caption{Graph of metric
potential versus $\emph{r}$.}
\end{figure}

\section{Physical Aspects of Compact Objects}

Here, we examine the physical features of compact objects by
graphical interpretation of fluid parameters, energy conditions,
stability analysis and speed of sound.

\subsection{Effective Fluid Parameters}

The fluid parameters should be maximum at the center. For this
reason, we consider small radii to analyze the smooth behavior of
compact stars. The plots of the effective matter variables are given
in Figure \textbf{4}, which shows positive behavior inside the
compact object and decreasing nature at the surface boundary. This
represents the compact behavior of the star at the center.
Furthermore, we examine that the gradient of effective fluid
parameters is negative which shows the high compactness of the
compact object. Here, we observe that stellar structures depend upon
the conserved quantities.
\begin{figure}
\epsfig{file=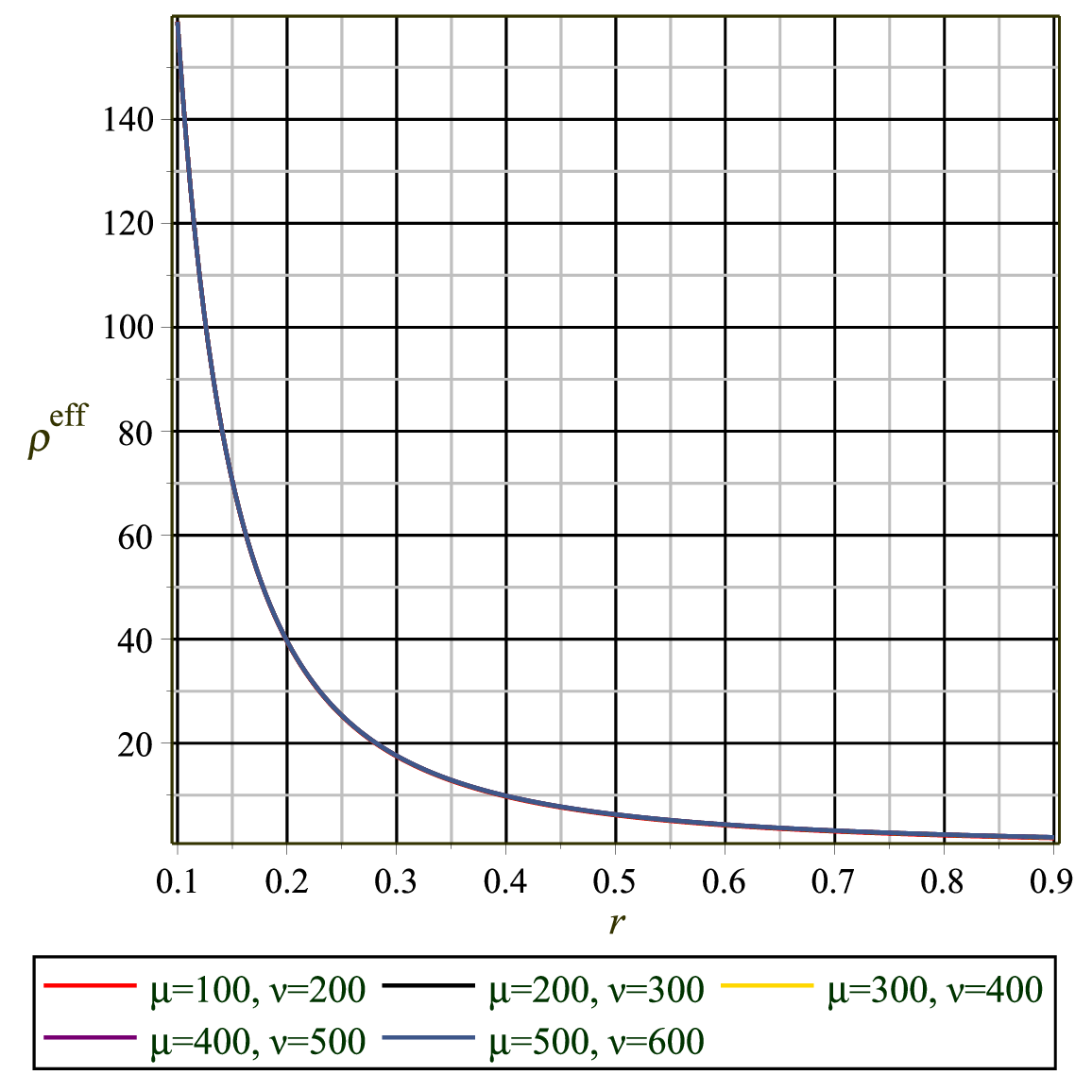,width=.5\linewidth}
\epsfig{file=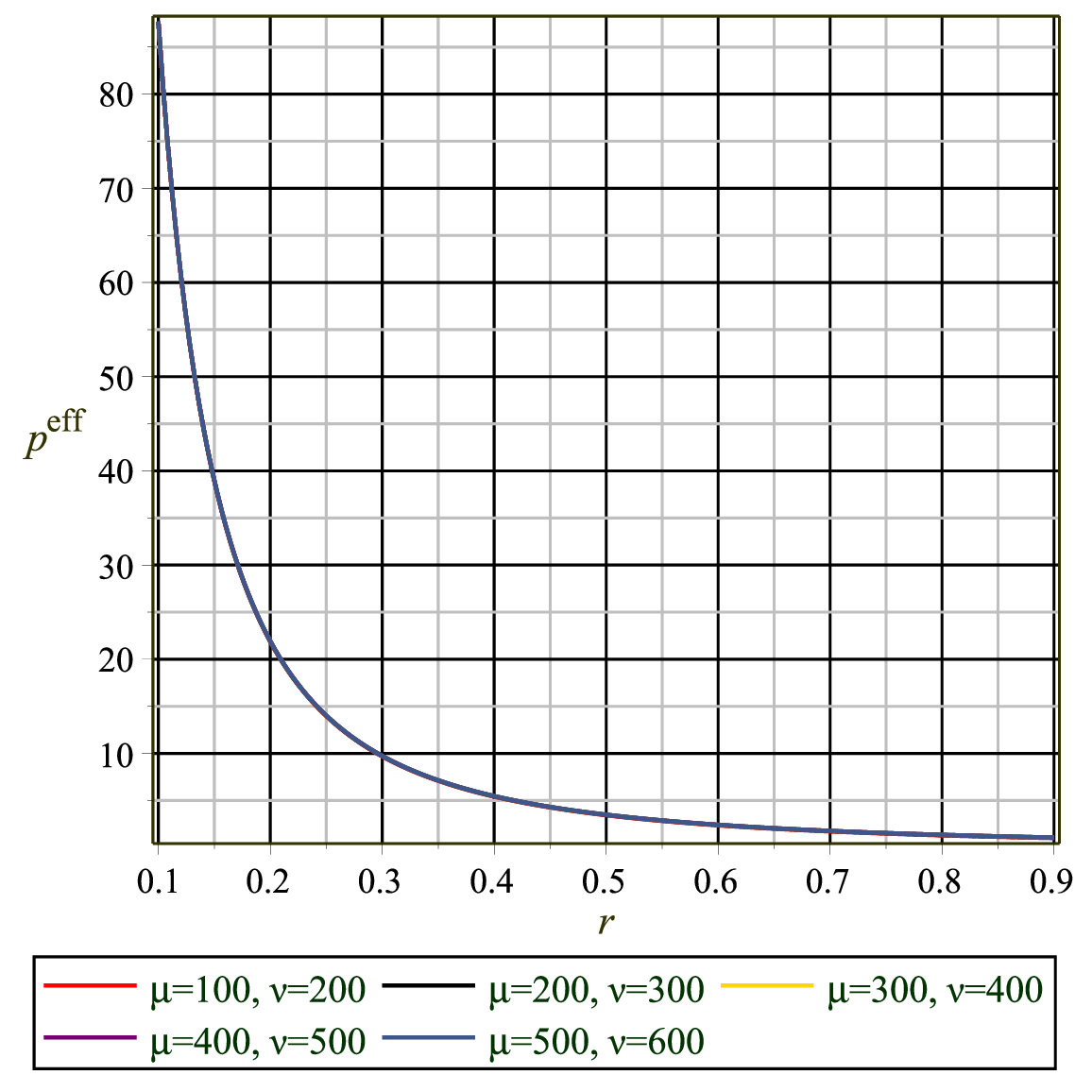,width=.5\linewidth} \caption{Behavior of
effective fluid parameters versus $\emph{r}$.}
\end{figure}

\subsection{Energy conditions}

The energy bounds are the key aspect in determining the physical
presence of cosmic geometries and viable fluid configuration. These
bounds must be satisfied for compact objects to have a physically
viable geometry. These bounds are the key aspects to investigate the
nature of the matter in the compact object determined as \cite{30}
\begin{eqnarray}\nonumber
&&\mathbb{NEC}:\quad \mathrm{\rho}^{eff}+\mathrm{p}^{eff}-\textit{A}\geq
0,\\\nonumber
&&\mathbb{WEC}:\quad\mathrm{\rho}^{eff}-\textit{A}\geq
0, \quad \mathrm{\rho}^{eff}+\mathrm{p}^{eff}-\textit{A}\geq 0,
\\\nonumber
&&\mathbb{SEC}:\quad\mathrm{\rho}^{eff}+\mathrm{p}^{eff}-\textit{A}\geq 0,
\quad \mathrm{\rho}^{eff}+3\mathrm{p}^{eff}-\textit{A}\geq 0,
\\\nonumber
&&\mathbb{DEC}:\quad\mathrm{\rho}^{eff}-\textit{A}\geq 0, \quad
\mathrm {\rho}^{eff}\pm \mathrm{p}^{eff}-\textit{A}\geq 0,
\end{eqnarray}
where $\mathbb{NEC}$, $\mathbb{WEC}$, $\mathbb{SEC}$ and
$\mathbb{DEC}$ define the null, weak, strong and dominant energy
conditions, respectively, while $\textit{A}
=\frac{1}{4e^{b}}\left(a'^{2}+2a''+4a'\emph{r}^{-1}-a'b'\right)$ is
an acceleration term that appears due to the matter source.

The energy density must be positive inside the compact star and
maximum at the center for the physically viable geometry. The
graphical behavior of energy bounds for various values of $\mu$ and
$\nu$ is shown in Figure \textbf{5}, which shows the viability and
consistency of our considered $f(\mathcal{R},\mathcal{T})$ model.
\begin{figure}
\epsfig{file=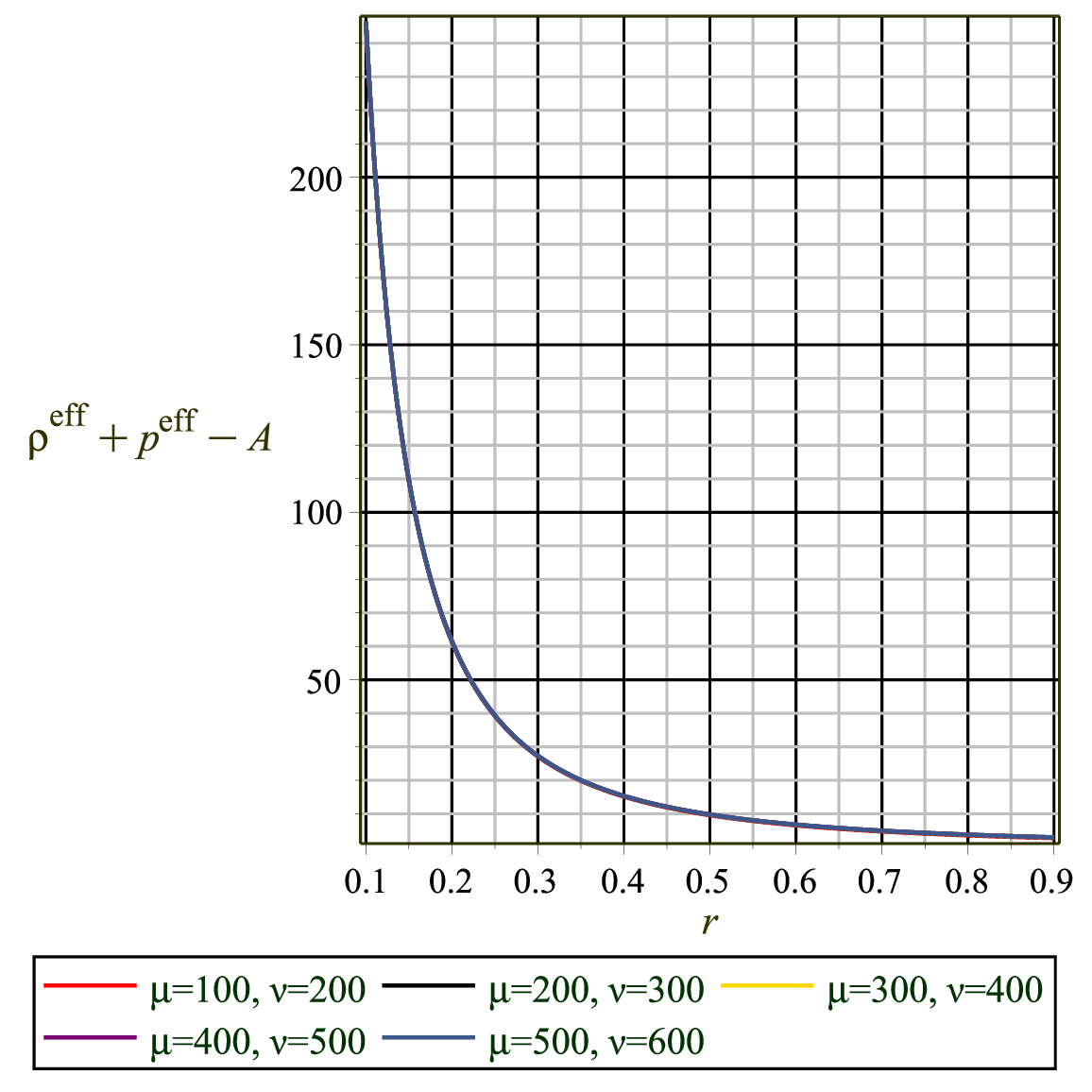,width=.5\linewidth}
\epsfig{file=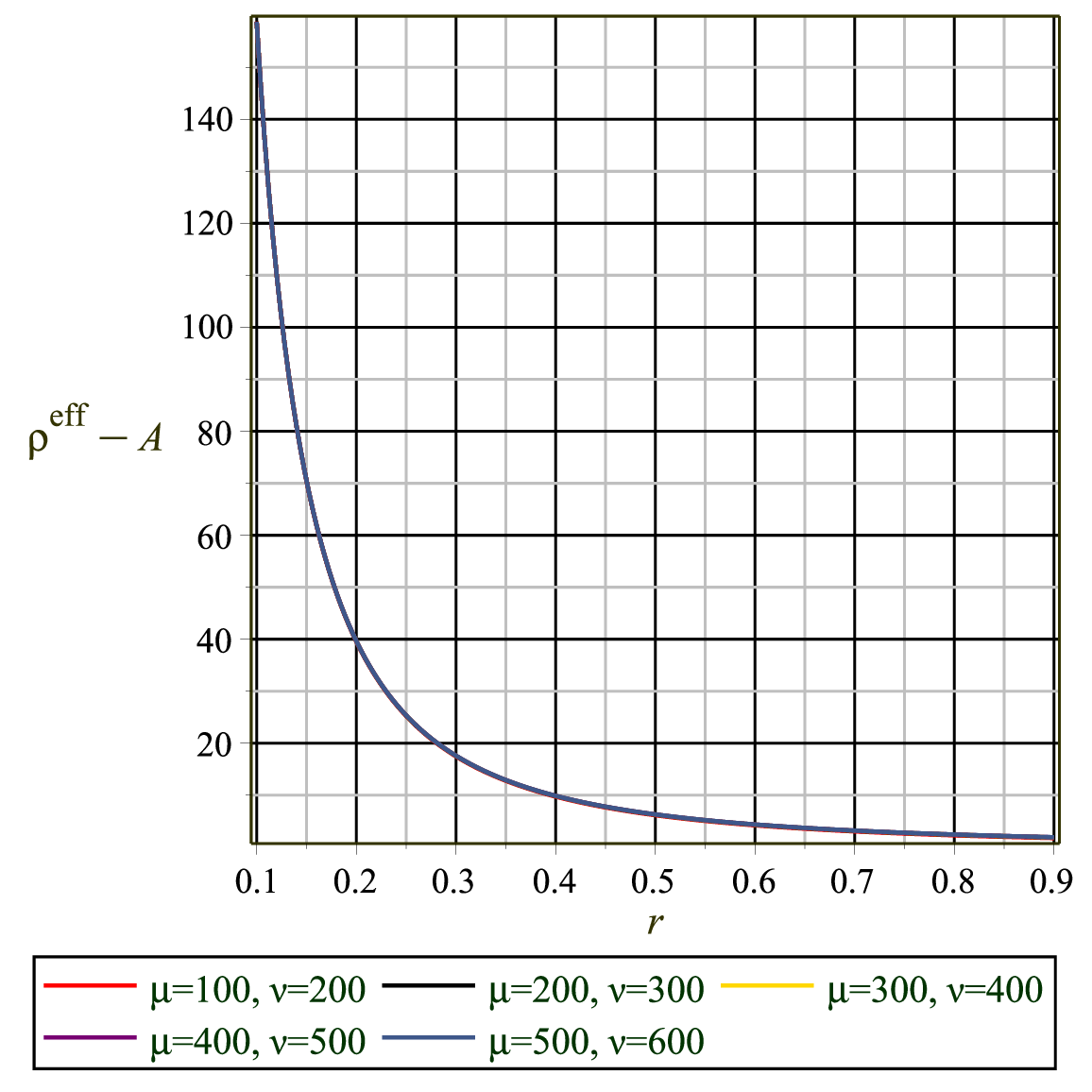,width=.5\linewidth}
\epsfig{file=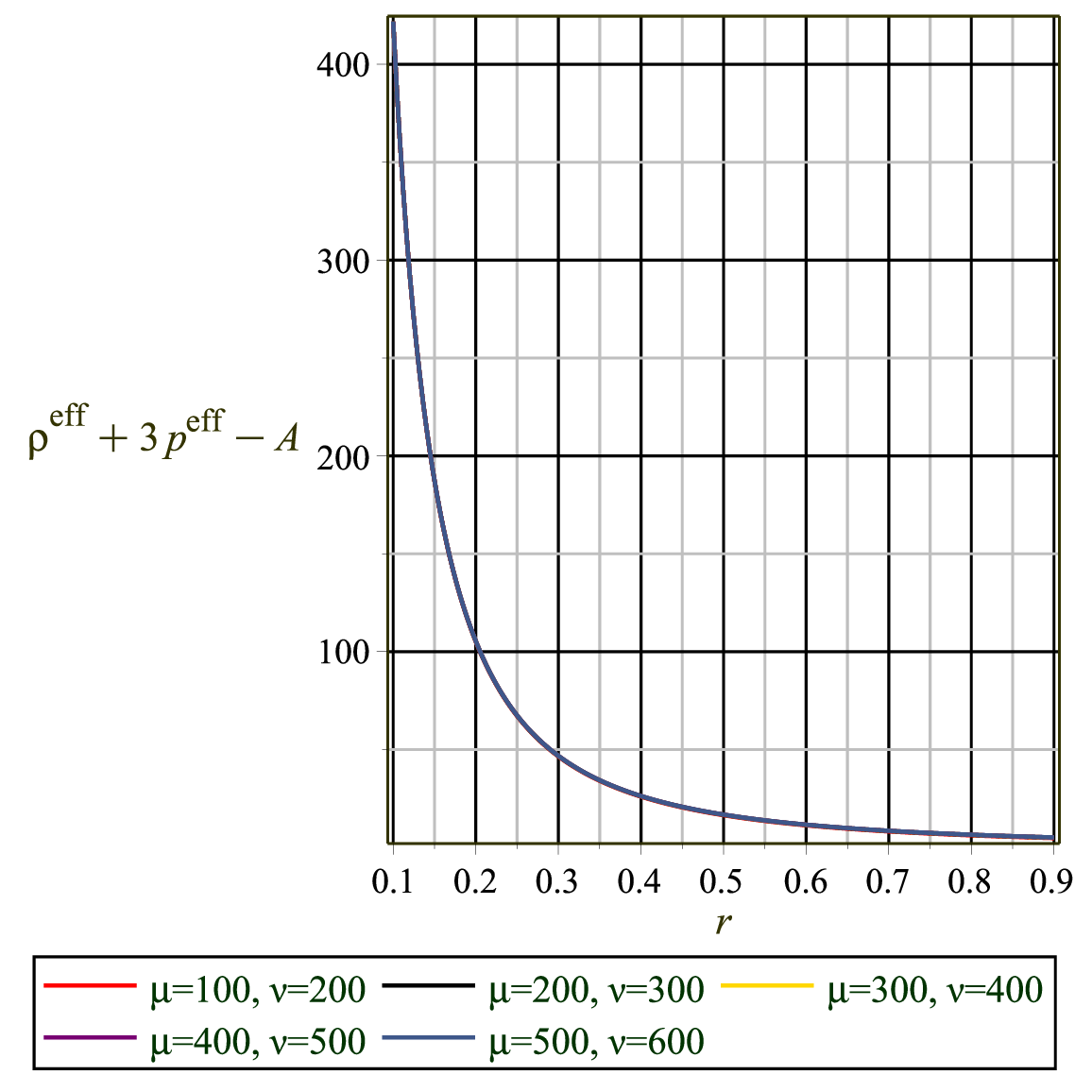,width=.5\linewidth}
\epsfig{file=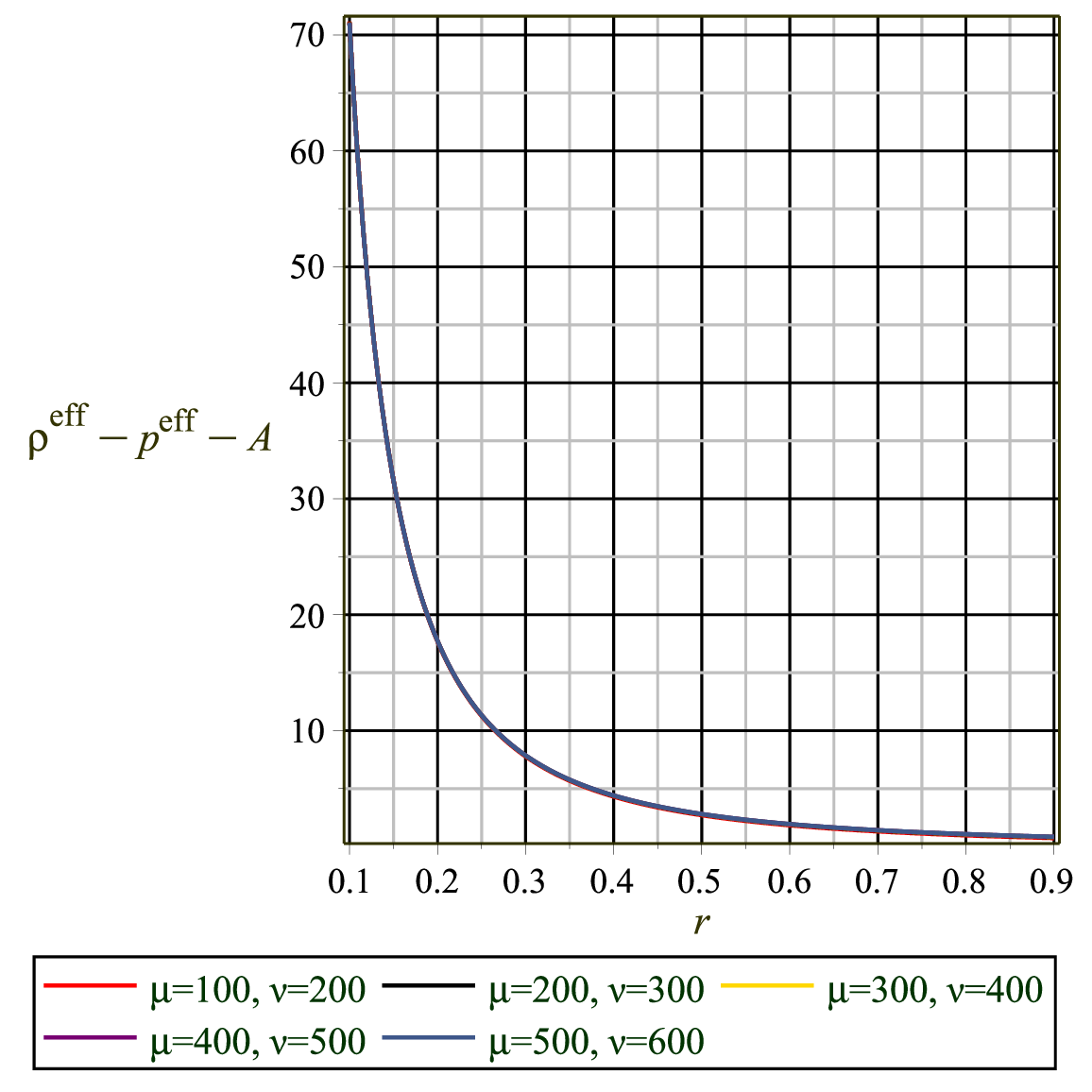,width=.5\linewidth} \caption{Behavior of energy
bounds.}
\end{figure}

\subsection{The Modified TOV Equation}

The conserved equation is expressed as
\begin{equation}\label{41}
\nabla^{\alpha}T^{eff}_{\alpha\beta}=0.
\end{equation}
We use a modified TOV equation with dust fluid distribution to
examine the equilibrium state of compact objects.
\begin{equation}\label{42}
-\frac{d\mathrm{p}^{eff}}{d\emph{r}}-\frac{a'}{2}
(\mathrm{\rho}^{eff}+\mathrm{p}^{eff})=0.
\end{equation}
This describes the relation of gravitational
$\left(\mathcal{F}_{\mathfrak{g}}\right)$ and hydrostatic
$\left(\mathcal{F}_{\mathfrak{h}}\right)$ force that describe the
stable state of stellar structure. In the view of Eq.(\ref{42}),
these forces can be divided as
$\mathcal{F}_{\mathfrak{g}}=-\frac{a'}{2}(\mathrm{\rho}^{eff}
+\mathrm{p}^{eff}) $ and
$\mathcal{F}_{\mathfrak{h}}=-\frac{d\mathrm{p} ^{eff}}{d\emph{r}}$.
The null impact of these forces $\left(\mathcal{F}_{
\mathfrak{h}}+\mathcal{F}_{\mathfrak{g}}= 0\right)$ determine the
presence of viable compact stars \cite{31}. The graphical behavior
of $\mathcal{F}_{\mathfrak{h}}$ and $\mathcal{F}_{\mathfrak{g}}$ for
different values of $\mu$ and $\nu$ is presented in Figure
\textbf{6}, which shows the equilibrium state of the stellar system.
\begin{figure}
\begin{center}
\epsfig{file=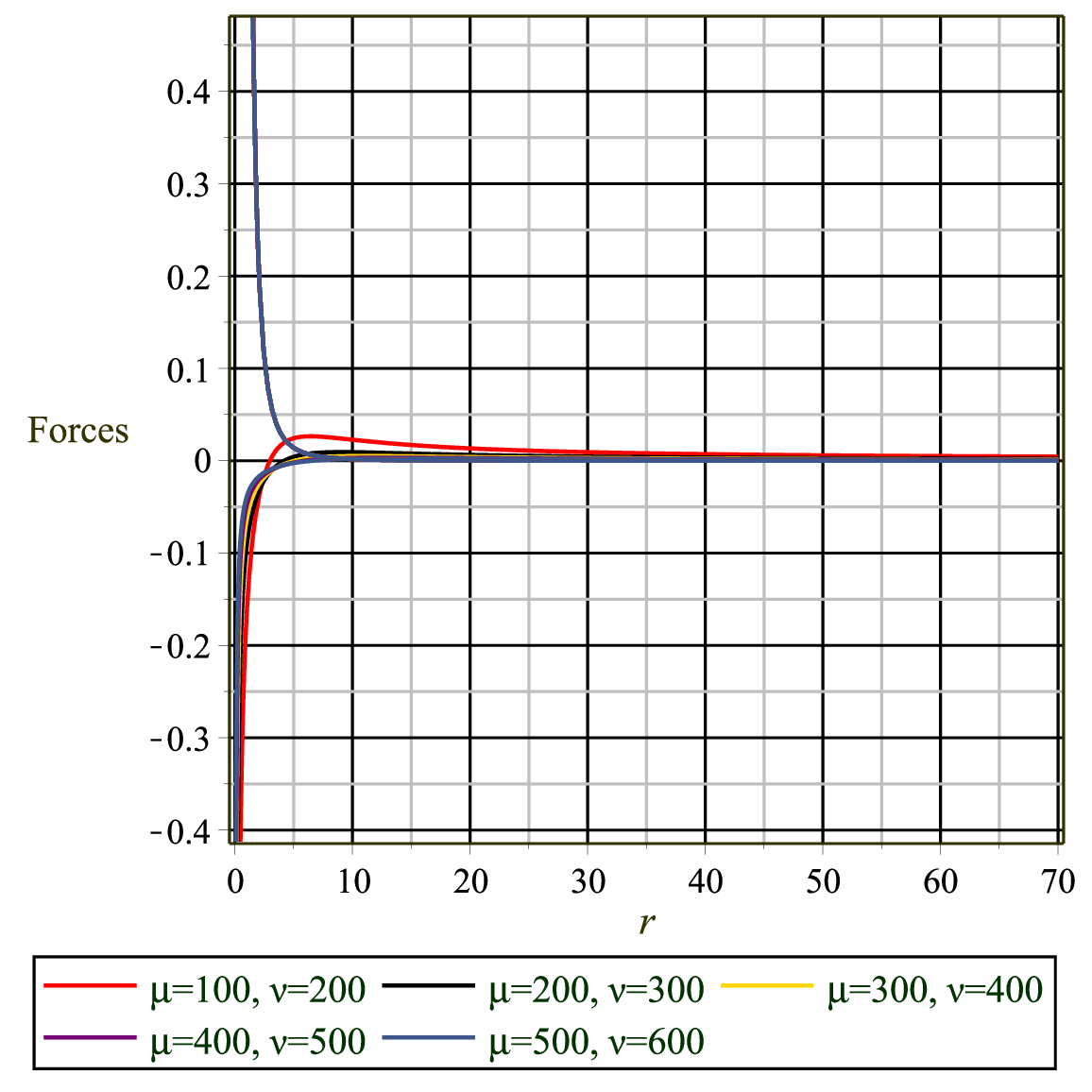,width=.55\linewidth} \caption{Behavior of
hydrostatic and gravitational forces ($\mathcal{F}_{\mathfrak{h}}$,
$\mathcal{F}_{\mathfrak{g}}$).}
\end{center}
\end{figure}

\subsection{Stability Analysis}

\subsubsection{Sound Speed}

The physical viability and consistency of a model are dependent on
the stability of compact objects. We assume Herrera's cracking
method \cite{32} to examine the stability of our proposed model.
This approach stated that sound speed $(v_{s} ^{2})$ must satisfy
the given condition $(0\leq v_{s}^{2}\leq1)$ and it is defined as
\begin{equation}\label{44}
v_{s}^{2}= \frac{d\mathrm{p}^{eff}}{d\mathrm{\rho}^{eff}}.
\end{equation}
Figure \textbf{7} shows that $v_{s}^{2}$ satisfies the required
condition and hence matter configuration is stable.
\begin{figure}
\begin{center}
\epsfig{file=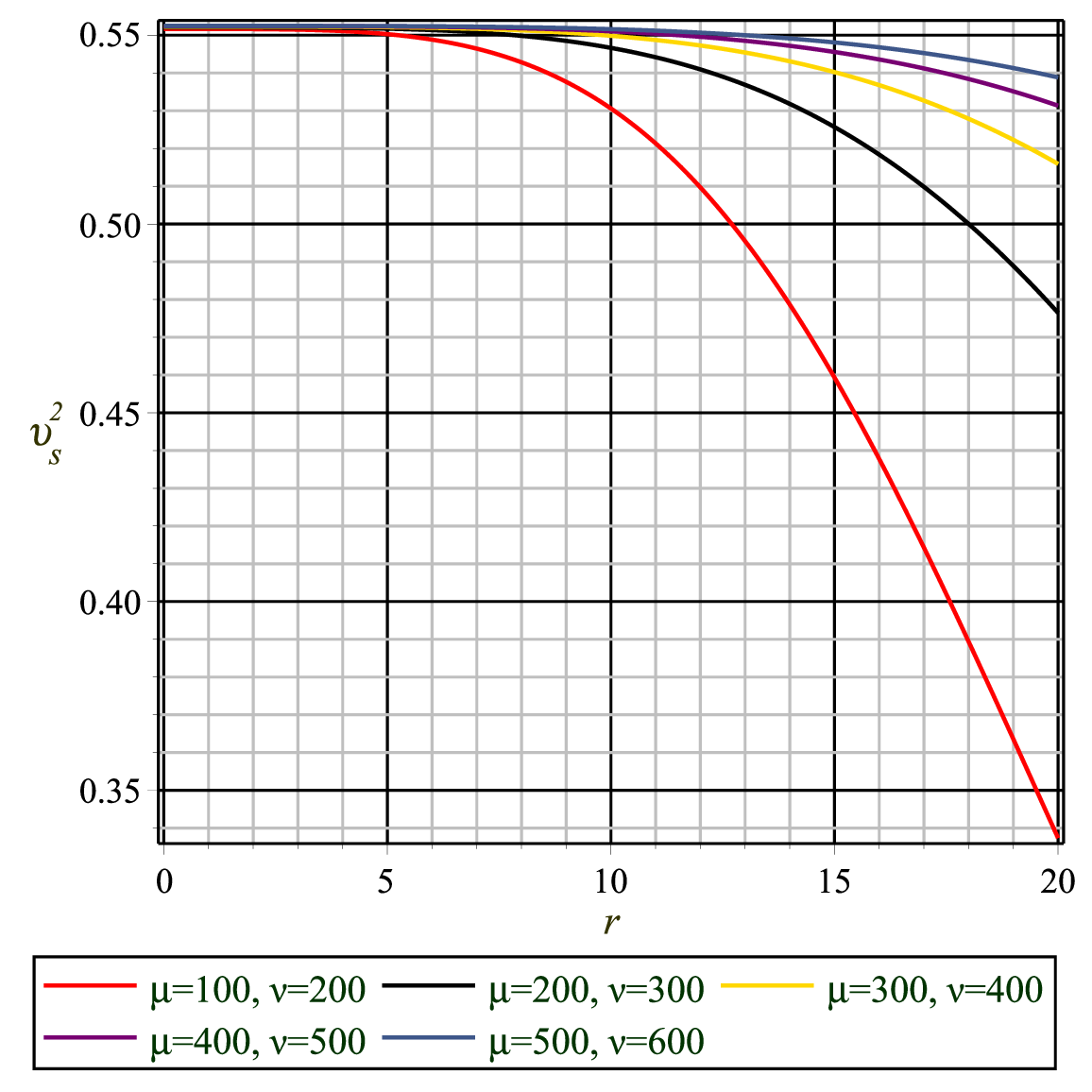,width=.55\linewidth} \caption{Variation of
speed of sound.}
\end{center}
\end{figure}

\subsubsection{Adiabatic Index}
Chandrasekhar \cite{32a} introduced the formalism to examine the
stability of the celestial object against radial perturbation. Many
researchers has been established and used this approach on the
astrophysical level \cite{32b}. The adiabatic index is defined as
\begin{equation}\nonumber
\Omega=\left(\frac{\mathrm{\rho}^{eff}+\mathrm{p}^{eff}}{\mathrm{p}^{eff}}\right)
\left(\frac{d\mathrm{p}^{eff}}{d\mathrm{\rho}^{eff}}\right).
\end{equation}
For a stable configuration, the adiabatic index should be greater
than 4/3 within the isotropic stellar system. The graphical behavior
of the adiabatic index is given in Figure \textbf{8}, which shows
that our system is in a stable state.
\begin{figure}\center
\epsfig{file=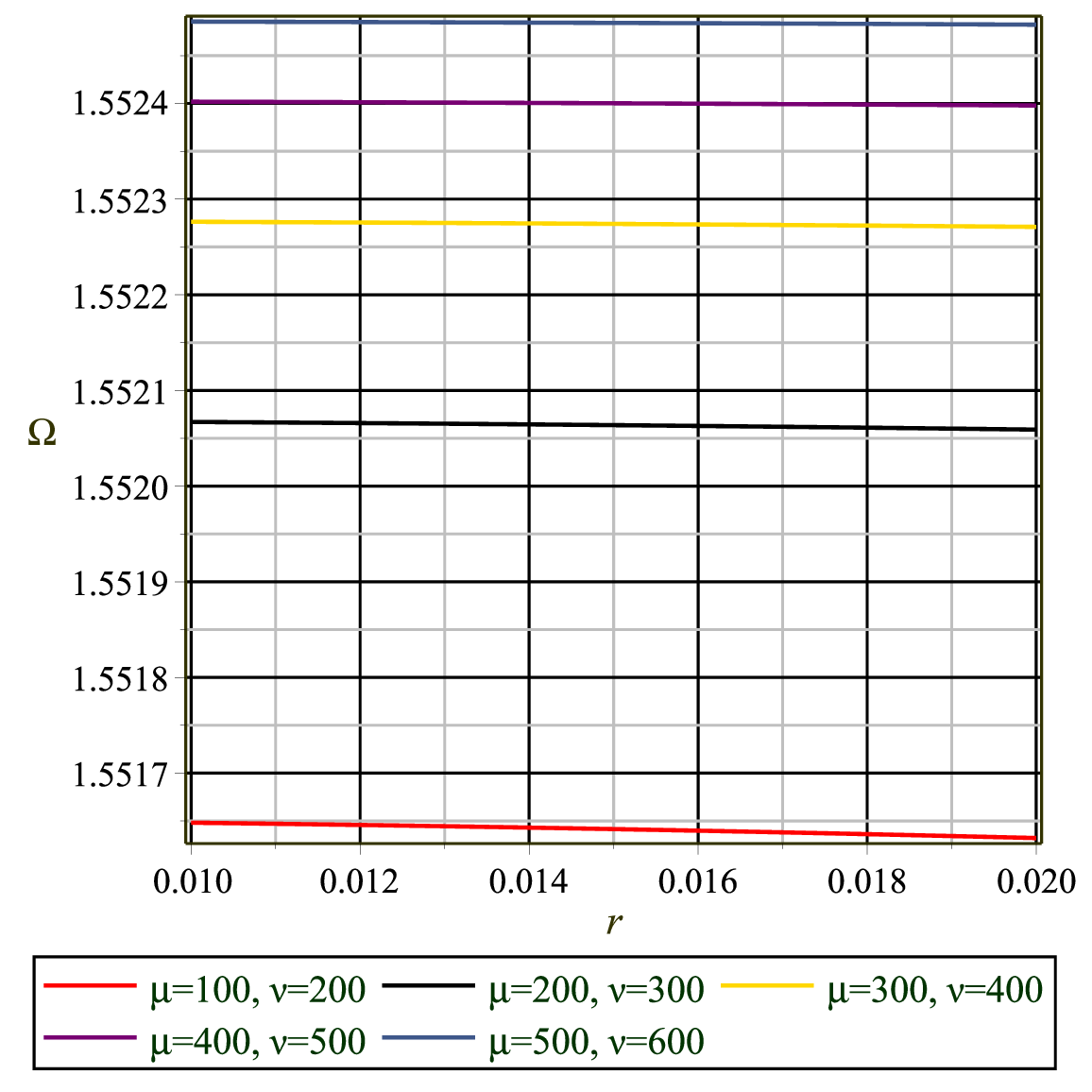,width=.5\linewidth} \caption{Graph of adiabatic
index corresponding to $\emph{r}$.}
\end{figure}

\subsection{Compactness and Surface Redshift}

The compactness factor is the ratio of mass and radius of a
celestial object. The mass of the star is directly proportional to
the radius as shown in Figure \textbf{9} and
$M(\emph{r})\rightarrow0$ as $\emph{r}\rightarrow0$, which
determines that the mass is an increasing function and regular at
the core of a star. The compactness factor is defined  in the
following form
\begin{equation}\nonumber
\gamma=\frac{M(\emph{r})}{\emph{r}}.
\end{equation}
The gravitational redshift plays a key role to examine the smooth
relation between particles in the stellar object. The gravitational
redshift is expressed as
\begin{equation}\nonumber
Z_{s}=\frac{1}{\sqrt{1-2\gamma}}-1.
\end{equation}
The graphical behavior of the compactness factor and surface
redshift is given in Figure \textbf{10}. These plots manifest that
the behavior of $\gamma$ and $Z_{s}$ is increasing as required.
\begin{figure}\center
\epsfig{file=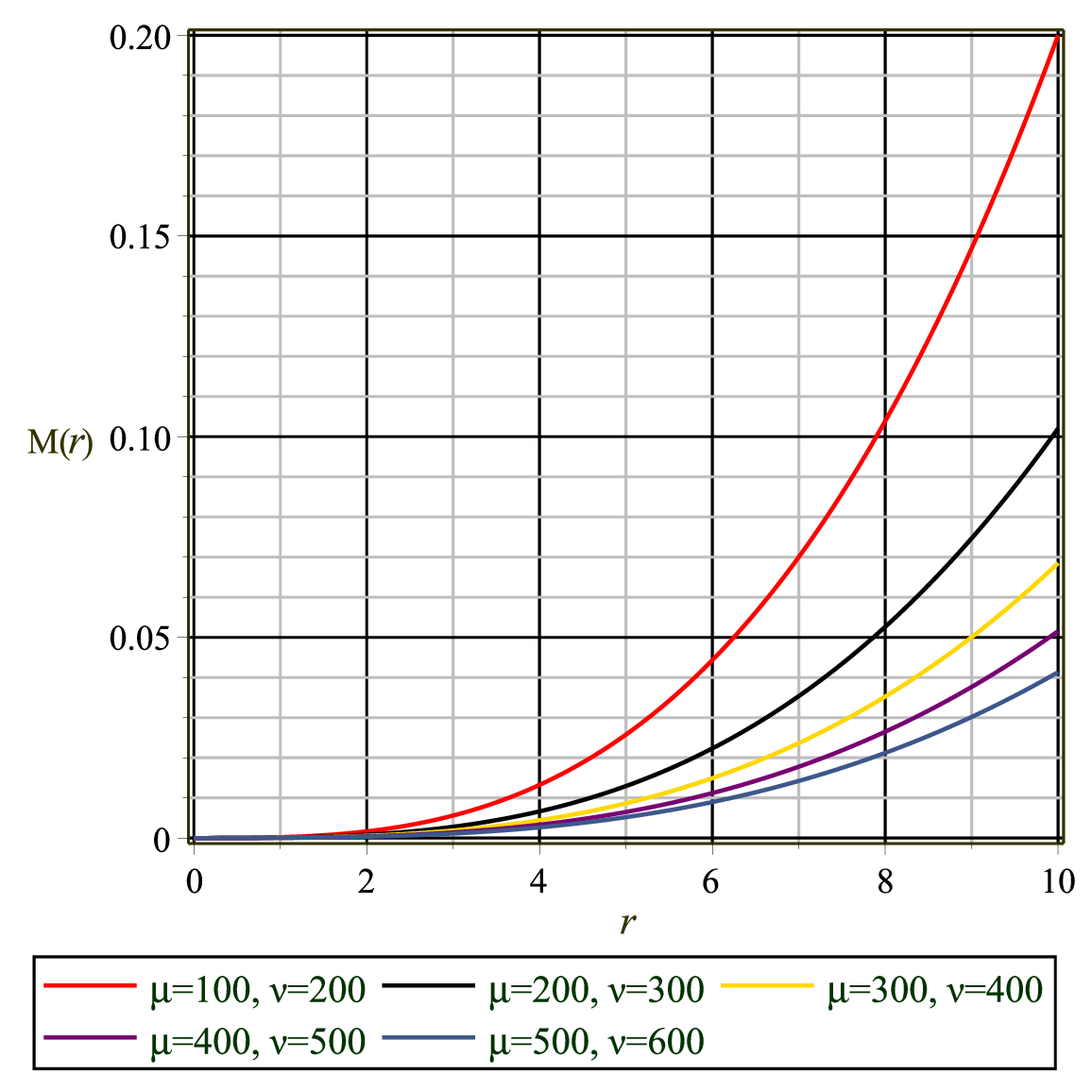,width=.5\linewidth} \caption{Behavior of the
mass function versus \emph{r}.}
\end{figure}
\begin{figure}
\epsfig{file=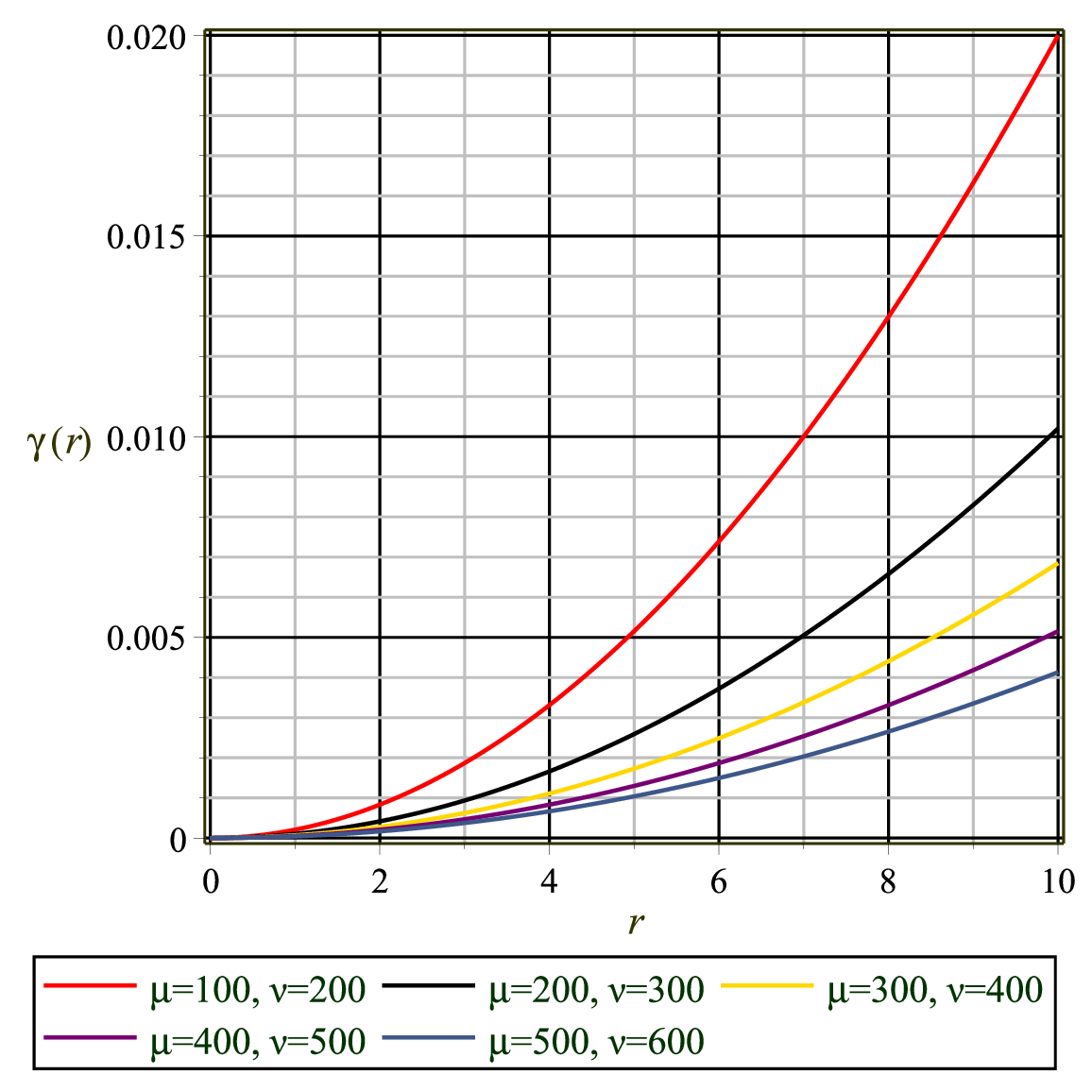,width=.5\linewidth}
\epsfig{file=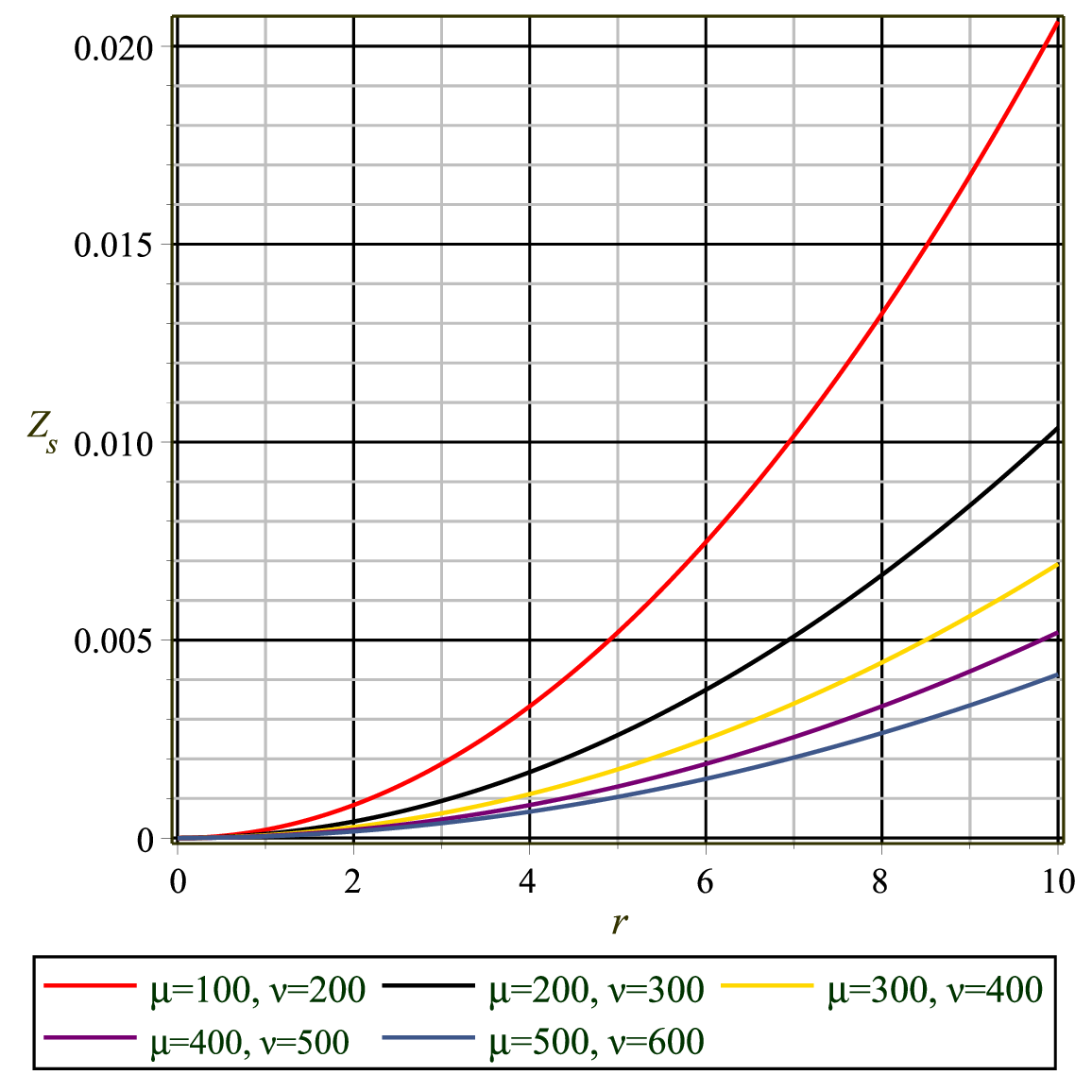,width=.5\linewidth} \caption{Graphs of the
compactness parameter and surface redshift versus \emph{r}.}
\end{figure}

\section{Concluding Remarks}

Noether symmetries are very useful to determine the exact solutions
of the physical system. The Lagrange multipliers reduce the complex
structure that eventually assists in evaluating the exact solutions.
In this article, we have examined the physical features of compact
stars through the NS technique. We have established Lagrangian of
this theory and formulated symmetry generators and corresponding
conserved parameters. The exact solutions of the field equations
have been analyzed for a specific model of this theory. The major
consequences of this paper are bestowed as follows.
\begin{itemize}
\item
The metric elements must be positive, finite and non-singular at
every point in the geometry of the star to obtain the physically
viable model. Figures \textbf{1}, \textbf{2} and \textbf{3} shows
the viability and the consistency of our metric elements.
\item
The effective fluid parameters should be maximum at the core of
compact objects. The graphs for small radius have been plotted to
show the smooth behavior of compact objects. These fluid parameters
have a maximum value at the center and decrease towards the boundary
that represents the viable behavior (Figure \textbf{4}).
\item
All energy constraints are well satisfied for a proposed model which
determine the viable matter (Figure \textbf{5}).
\item
We have examined the equilibrium state (Figure \textbf{6}) through
TOV equation and found that our model is compatible with stability
condition (Figures \textbf{7} and \textbf{8}).
\item
The graphical analysis of the mass function is found to be
increasing and regular at the center of the star (Figure
\textbf{9}).
\item
Finally, we have found (Figure \textbf{10}) that the behavior of
compactness parameter and gravitational redshift function is
increasing as required.
\end{itemize}
We have concluded that compact objects in this theory through the NS
technique depend on the conserved and model parameters. We have
investigated that all physical attributes of stellar objects fulfill
the physically viable pattern. We found that the NS technique in
$f(\mathcal{R},\mathcal{T})$ gravity yields a more realistic and
stable model.

\end{document}